\documentclass{article} 
\usepackage{graphicx,xspace}
\usepackage[acronym,xindy,toc,nonumberlist,shortcuts]{glossaries} 
\usepackage{url}

\usepackage{floatrow}
\usepackage{enumitem}
\usepackage{graphics}
\usepackage{array}
\usepackage[table]{xcolor}
\newcolumntype{P}[1]{>{\raggedright\arraybackslash}p{#1}}

\usepackage{natbib}

\newglossaryentry{v}{name=carto-vandalism}
\newglossaryentry{c}{name=crowdsourced cartography}
\newglossaryentry{vgi}{name=volunteered geographic information}
\newglossaryentry{cpted}{name=crime prevention through environmental design}
\newacronym{lod}{LOD}{linked open data}

\newglossaryentry{hci}{name=human-computer interaction}
\newacronym{gir}{GIR}{geographic information retrieval}
\newacronym{ai}{AI}{artificial intelligence}
\newacronym{ir}{IR}{information retrieval}
\newacronym{gis}{GIS}{geographic information system}
\newacronym{idf}{IDF}{Inverse Document Frequency}
\newacronym{lsa}{LSA}{Latent Semantic Analysis}
\newacronym{poipl}{POI}{points of interest}
\newacronym{poi}{POI}{point of interest}
\newacronym{vsm}{VSM}{vector space model}
\newacronym{pos}{POS}{part-of-speech}
\newacronym{gwap}{GWAP}{game with a purpose}
\newacronym{www}{WWW}{World Wide Web}
\newacronym{skos}{SKOS}{W3C Simple Knowledge Organization System}
\newacronym{rdf}{RDF}{Resource Description Framework}
\newacronym{owl}{OWL}{Web Ontology Language}
\newacronym{oaei}{OAEI}{Ontology Alignment Evaluation Initiative}
\newacronym{giscience}{GIScience}{geographic information science}
\newacronym{wsddef}{WSD}{word sense disambiguation}
\newacronym{dl}{DL}{description logic}
\newacronym{mdsmsim}{MDSM}{Matching-Distance Similarity Measure}
\newacronym{oursurvey}{GeReSiD}{Geo Relatedness and Similarity Dataset}
\newacronym{ira}{IRA}{interrater agreement}

\newacronym{irr}{IRR}{interrater reliability}
\newacronym{mds}{MDS}{multidimensional scaling}
\newacronym{lbs}{LBS}{location-based services}
\newacronym{oss}{FOSS}{free and open-source software}
\newacronym{sdts}{SDTS}{Spatial Data Transfer Standard}
\newacronym{tfidf}{TF-IDF}{Term Frequency-Inverse Document Frequency}
\newacronym{ogc}{OGC}{Open Geospatial Consortium}

\newglossaryentry{w2}{name=Web 2.0,
	description={TODO}
}

\newglossaryentry{bow}{name=bag-of-words}
\newglossaryentry{osm}{name=Open\-Street\-Map}
\newglossaryentry{wm}{name=Wiki\-Mapia}

\newglossaryentry{algo}{name=\emph{Voc2WordNet}}
\newglossaryentry{algoplain}{name=Voc2WordNet}

\newglossaryentry{lodcloud}{name=LOD cloud}

\newglossaryentry{stratag}{name=Strategic Research in Advanced Geotechnologies (StratAG)}

\newglossaryentry{wn}{name=WordNet}

\newglossaryentry{kb}{name={knowledge base}}
\newglossaryentry{gkb}{name={geo-knowledge base}}

\newglossaryentry{nuim}{name={National University of Ireland, Maynooth},
	description={TODO}
}

\newglossaryentry{ucd}{name={University College Dublin},
	description={TODO}
}

\newglossaryentry{osmsim}{name=\textsc{OSM-TagSim},
	description={TODO}
}

\newglossaryentry{lexsim}{name=\textsc{osm-sim_{lex}},
	description={TODO}
}

\newglossaryentry{simdl}{name=Sim-DL,
	description={TODO}
}

\newglossaryentry{netsim}{name=\textsc{osm-sim_{sim}},
	description={TODO}
}

\newglossaryentry{dbp}{name=DBpedia,
	description={TODO}
}

\newglossaryentry{lgd}{name=LinkedGeoData,
	description={TODO}
}

\newglossaryentry{sgw}{name=Semantic Geospatial Web,
	description={TODO}
}

\newglossaryentry{sw}{name=Semantic Web,
	description={TODO}
}

\newglossaryentry{webplatform}{name=Web platform for map personalisation and visualisation,
	description={TODO}
}

\newglossaryentry{wsd}{name=word sense disambiguation,
	description={TODO}
}

\newglossaryentry{os}{name=open source,
	description={TODO}
}

\newglossaryentry{osn}{name=OSM Semantic Network,
	description={TODO}
}

\newglossaryentry{nlp}{name=natural language processing,
	description={TODO}
}

\newglossaryentry{owc}{name=OSM Wiki Crawler,
	description={TODO}
}

\newglossaryentry{oww}{name=OSM Wiki website,
	description={\url{http://wiki.openstreetmap.org}}
}

\newglossaryentry{mdsm}{name=MDSM evaluation dataset,
	description={TODO}
}

\newcommand{\footurl}[1]{\footnote{\url{#1}}}

\title{Defacing the map: Cartographic vandalism in the digital commons}
\date{\small\textit{Author copy. Cartographic Journal, 2014.}}

\author{
Andrea Ballatore\\
Department of Computer Science\\
National University of Ireland, Maynooth\\
\texttt{andrea.ballatore@nuim.ie}
}
 
\begin{document}
\maketitle

\begin{abstract}

\noindent This article addresses the emergent phenomenon of \emph{\gls{v}}, the intentional defacement of collaborative cartographic digital artefacts in the context of  \gls{vgi}.
Through a qualitative analysis of reported incidents in \gls{wm} and \gls{osm}, a typology  of this kind of vandalism is outlined, including
play,  
ideological, 
fantasy, 
artistic, and  
industrial \gls{v}, as well as carto-spam.
Two families of counter-stra\-te\-gies deployed in amateur mapping communities are discussed.
First, the contributors organise forms of policing, based on volunteered community involvement, patrolling the maps and reporting incidents.
Second, the detection of \gls{v} can be supported by automated tools, based either on explicit rules or on machine learning.

\end{abstract}
\vspace{1em}

\newpage

\section{Introduction}
\label{sec:intro}
\glsresetall

Maps are widely recognised as powerful and versatile representations of geographic realities, loaded with cultural, aesthetic, and practical meanings.
Thanks to ubiquitous and inexpensive personal computers, smartphones, and location technologies, mapping practices are currently experiencing a reconfiguration, driven by online modes of production, delivery, and consumption, generating informational commons of reusable geographic information.
These forms of \gls{c} inherit well-established mapping conventions and traditions, but also present novel aspects that deserve attention \citep{dodge:2013:mappingexp}.
Collaborative online mapping relies on commons-based peer production, almost entirely mediated by digital communication tools.
As \cite{benkler:2006:commons} put it, commons-based peer production is a socio-technical system that relies on ``collaboration among large groups of individuals \ldots who cooperate effectively to provide information, knowledge, or cultural goods without relying on either market pricing or managerial hierarchies to coordinate their common enterprise'' (p. 394).

In the geographic domain, commons-based peer production has transformed the traditional expert-driven models that had dominated the field.
\cite{Goodchild:2007:citizens} used the term \emph{volunteered geographic information} to describe the complex constellation of amateur mapping projects,
while \cite{graham:2010:neogeography} identified in these phenomena a common drive to create virtual versions of places and geographic realities. 
In such projects, contributors form self-regulating communities to create, maintain, and promote the usage of digital geospatial artefacts.
Wikified maps, free gazetteers, mash-ups, and open geo-knowledge bases create an inter-linked ecosystem of geospatial commons \citep{Ballatore:2012:survey}.
In addition, these large collaborations form a communal space in which individuals can engage with their geographic surroundings in novel ways \citep{elwood:2008:volunteered}.

Among others, a major difference exists between traditional and commons-based peer cartography.
Traditional cartographic production is conducted by experts in a private environment, which limits the possibility of malicious actions on the data.
This is manifestly not the case in commons-based peer production projects such as \gls{wm} and \gls{osm}.
This article investigates \emph{\gls{v}}, an emergent form of deviant behaviour in collaborative production environments directed at geographic information.
Unlike online open datasets, physical geospatial artefacts can be vandalised only when exposed in public spaces, such as in the case of tourist maps and cartographic monuments.
A notable instance of such vandalism can be found in the history of modern Italy, when in 1943 one of Mussolini's `imperial maps' in Rome was defaced with red paint, symbolically attacking Fascism's geographic expansionism \citep{minor:1999:mussolini}.\footnote{In 1934, Benito Mussolini commissioned and supervised the construction of a monument consisting of five `imperial maps,' overlooking the \emph{Via dei Fori Imperiali} in Rome.
The first four tablets represented distinct phases of the expansion of the Roman Empire, while the fifth showed the borders of Fascist Italy, inviting a comparison between the Roman Empire and his own regime.} 

Mutatis mutandis, the digital maps that are generated today through com\-mons-based peer production suffer from a variety of forms of vandalism. 
Because \gls{c} is intrinsically open to anonymous contributions, the core geospatial artefacts are highly vulnerable to malicious contributions.
Commons-based peer production relies heavily on implicit trust in contributors, and on the goodwill of participants to construct value rather than destroy it.
For example, anybody can create an account and edit the vector map in \gls{osm}, the most prominent peer production cartographic initiative, and see their changes immediately visible online.
Whilst malicious edits on Wiki\-pedia have been the object of much research, \gls{v}, i.e. vandalism aimed at geographic artefacts, is a little-understood phenomenon.
In the recent debate on the quality of crowdsourced geographic information, this kind of vandalism is often mentioned, but has not been analysed organically \citep{Flanagin:2008:credibility,mooney:2010:towards,goodchild:2012:assuring}.

The perception of \gls{v} plays a crucial role in the credibility -- or lack thereof -- of \gls{c} \citep{Flanagin:2008:credibility}.
Therefore the stakes surrounding it are high.
Unsurprisingly, \gls{c} advocates tend to minimise the severity and importance of \gls{v}.
\gls{osm}'s founder Steve \cite{Coast:2010:bestmap} argued that ``malicious edits are probably the least significant and will always exist'' (para. 43).
By contrast, commercial providers of geographic information cite vandalism as a serious flaw of \gls{c}.
In a promotional newsletter, the company \cite{tomtom:2012:opensource} claims that user-generated maps are ``wide open to attack'' and therefore unfit for critical applications such as routing, generating particular anxiety by evoking the possibility that anybody can change the direction of one-way streets, with potentially dire consequences for drivers.
Although both claims bear some degree of truth, they are influenced by vested interests, and a deeper understanding of the reasons behind \gls{v} is necessary.

The phenomenon of \gls{v} can be observed from several, complementary perspectives.
\Gls{vgi} is generated by online communities, through forms of network sociality \citep{wittel:2001:netsociality}, 
and the weak social bonds that dominate online communities can indeed favour deviant behaviour.\footnote{For a discussion of the problematic notion of `online community' see Williams (2006), pp. 14-17. }
Definitional problems surround the emotionally-loaded term `vandalism,' used to label a variety of behaviours whose boundaries are notoriously difficult to define and whose targets include buildings, vehicles, as well as digital artefacts such as websites \citep{goldstein:1996:psychology,williams:2007:policing}.
From a legal viewpoint, unlike other types of `serious' cybercrime, such as the trade of child pornography or credit card cloning, the vandalisation of digital commons constitutes an uncharted gray area. 

This does not imply that \gls{v} has no tangible consequences for projects and contributors.
When vandalised, digital artefacts see their use value being diminished, and their appeal reduced.
As a result, communities have to devote part of their scarce resources to policing and other anti-vandalism activities that are often dull and unrewarding for contributors.
The possible causes and consequences of \gls{v} need be discussed in the framework of the unique properties of \gls{c}, emphasising the peculiarities that distinguish geospatial artefacts from other forms of digital commons.

The remainder of this article sheds light on these issues as follows.
Section \ref{sec:vanda} discusses the continuities of digital \gls{v} with physical vandalism, which has been studied from sociological and psychological standpoints.
Based on a qualitative analysis of incidents reported in \gls{osm} and \gls{wm}, two prominent \gls{vgi} projects, Section \ref{sec:classif} proposes a classification of the multiple facets of \gls{v}.
Amateur mapping communities need strategies to defend themselves from \gls{v}, and Section \ref{sec:counterv} surveys currently adopted counter-strategies.
Finally, Section \ref{sec:concl} draws conclusions, pointing out directions for future research about \gls{v}.

\section{Vandalism, physical and digital}
\label{sec:vanda}

The term `vandalism' refers to a variety of socially constructed phenomena, and no clear academic consensus has been established about its scope.
Reviewing its many definitions, \cite{moser:1992:vandalism} suggested that vandalism is ``a hodgepodge concept that covers behavior for which motivations are extremely different'' (p. 51).
The definitions of vandalism differ as they take into account the caused damage, the motivation of the human actor, and/or the context of the incident.
Vandalism is a ubiquitous and visible social phenomenon, in which intentional damage was performed on a variety of objects, including buildings, public toilets, vehicles, furniture, infrastructures, and works of art such as paintings, monuments, and sculptures.
Most vandalism incidents are far from being random, senseless acts, and several competing theories to explain its reasons have been proposed.
A consensus exists around the general meaningfulness of vandalism as a form of social communication between the offender and an imagined or real audience.
The next sections review theories of vandalism directed towards physical and digital objects, and discuss the issue of deviance in \gls{vgi} communities. 

\subsection{Physical vandalism}

Studies on vandalism against physical objects have been conducted mostly in the broad framework of criminology, harnessing theoretical tools from social psychology and sociology, particularly in the 1970s and 1980s.
The key preoccupation of these social science researchers is the reduction of such deviant behaviours, and their societal and economic negative consequences \citep{goldstein:1996:psychology}. 
South African criminologist Stanley \cite{cohen:1973:property} outlined a broad and influential classification of types of vandalism, based on the offender's purpose. 
His classification includes the following categories:
\emph{acquisitive vandalism} (theft or looting);
\emph{tactical/ideological vandalism} (to attract attention around a political or social issue); 
\emph{vindictive vandalism} (for revenge against somebody);
\emph{play vandalism} (unintentional damage resulting from children's games); and
\emph{malicious vandalism} (violent outpouring of rage).

In the same decade, \cite{allen:1978:aesthetic} proposed an aesthetic theory of vandalism, focusing on its pleasure-arousing potential, which depends on the specific traits of the target objects.
By defacing objects, offenders tend to alter their structure towards simpler configurations, and therefore objects of complex design are more likely to be vandalised than simple ones.
\cite{fisher:1982:equity} have argued that perceived social inequality is a key cause of vandalism, which is perpetrated as an iniquity resolution mechanism.
Their model accounts in particular for vandalism directed at public property, conceptualised by offenders as a symbol of an unfair status quo.  
Furthermore, \cite{sutton:1987:differential} expanded Cohen's classification, identifying a new type of vandalism, i.e. peer status motivated vandalism.
Mostly conducted by groups rather than individuals, this type of vandalism is perpetrated to gain or maintain peer status, for example damaging a vehicle on a dare.

From a more conservative standpoint, \cite{kelling:1982:brokenwindow} outlined their widely discussed `broken window theory,' claiming that minor crime such as vandalism affects the environment, and the resulting physical disorder tends to generate more serious crime.
The linkage between low and high intensity deviance has been also stressed by \cite{goldstein:1996:psychology} as a key motivation to counter vandalism.
However, these theories do not seem to capture the aforementioned essential drives behind physical vandalism, and have been challenged as inadequate to explain the socio-structural conditions that cause crime \citep{gau:2010:revbrokenwindow}. 
Over the last two decades, attempts to counter physical vandalism have been predominantly conducted in the framework of `\gls{cpted},' applying architectural design patterns to discourage deviant behaviour \citep{cozens:2008:histcpted}.

\subsection{Digital vandalism}


With the emergence of online spaces in the 1990s, vandalism found new avenues of expression, as happened with other types of crime that soon appeared on the Web.
The most common form of cybercrime is the defacement of websites, often for satirical or playful purposes \citep{furnell:2002:cybercrime}.
\cite{williams:2004:understandingking} defined the term `online vandalism' to refer to deviant behaviour in online communities, particularly the defacement of digital artifacts built collaboratively, often as part of cyber-bullying.
The environmental structure of online spaces determines specific, peculiar aspects of online vandalism.
The so-called `online disinhibition effect' consists of a reduction in social inhibitions and constraints, fostered by the perceived anonymity, invisibility, asynchronicity, dissociative imagination, and lack of authority in online spaces \citep{suler:2004:disinhibition}.

However, the many continuities between online digital vandalism and offline physical vandalism should not be understated, it is important to avoid the trap of what \cite{jurgenson:2012:atomsbits} named `digital dualism,' i.e. the tendency of overlooking the substantial intermesh between the online and offline.
Similarly to physical vandalism, `digital vandalism' generates a large number of low-impact incidents, is rarely reported, prosecuted and punished, and is -- quite reasonably -- perceived as a less serious threat than phishing, identity theft, and other obviously harmful activities.
The inherent openness of commons-based peer production projects makes them ideal targets for online vandalism. 

Because of its global success and visibility, vandalism against Wikipedia is the subject of an active research area, thoroughly surveyed by \cite{nielsen:2012:wikipediasurvey}.
The open infrastructure of Wikipedia is subject to structural vandalism, i.e. damage that is regularly inflicted on the digital artefact at the core of the community efforts.
The surprising success of Wikipedia relies on its ability keep such structural vandalism at bay, rapidly reverting suspicious edits.
Users deliberately delete valid information or enter incorrect facts in Wikipedia with a variety of playful or malicious purposes.

Because Wikipedia, as a form of digital commons, does not host information of high economic value, such as credit card numbers, authorities tend not to devote resources to counter online vandalism.
As a result, offenders have high rewards in terms of visibility and pleasure-arousing effects, and virtually no risk, a highly criminogenic combination.
Predictably, articles about controversial topics (e.g. abortion, divisive politicians, sexuality, contested borders) tend to be vandalised more often than others, and
correct but damaging information is often removed directly by affected individuals and organisations.
Notable cases of Wikipedia vandalism include obituaries of living celebrities, humorously incongruous facts, obscenities, and political propaganda. 


\subsection{Deviance in cartographic communities}
\label{sec:deviancevgi}

\Gls{c} is sustained by complex social networks, in which individuals deploy an inter-dependent combination of physical objects (personal computers, servers, cloud computing facilities, GPS sensors, and smartphones) and digital artefacts (datasets, websites, documentation, and software tools).  
The purpose of such \emph{ad hoc} communities is the development and the maintenance  digital artefacts, such as datasets, websites, and databases.
Community members co-operate through a variety of social media, including mailing lists, forums, private e-mails, wikis, Web conferences and, to a limited extent, through face-to-face interaction such as mapping parties.
These communities are formed around an explicit and shared purpose, coordinating and sustaining inter-personal bonds through a combination of online and offline communication.
As in any human community, behavioural norms are established, and deviance from these norms is discouraged and sanctioned with a variety of incentives and punishments \citep{williams:2006:virtualcriminal}.


While the vast majority of literature analyses \gls{vgi} data -- particularly \gls{osm} \cite[e.g.][]{haklay:2010:good,mooney:2012:characteristics} -- fewer studies directly observe the underlying communities.
Amateur mapping communities are characterised by social, technological and geographic divides.
Human geographers study \gls{vgi} as a social practice, focusing particularly on the inequalities between people and places involved in the generation of knowledge \citep{elwood:2012:researching}. 
Another dimension of study is that of the scope, motivations, and conditions of commons-based peer production of maps.

In these communities, social ties are less stable than in local communities, and the emotional attachment to the digital artefacts plays a crucial role in preserving the community from disaggregation.
Digital cartographic artefacts, such as the \gls{osm} world map, constitute the barycentre of the community, whose boundaries can shift considerably over short periods of time, with many new members joining and current members leaving.
\cite{coleman:2009:volunteered} have investigated the motivation of contributors, classifying them on a spectrum ranging from neophyte to expert authority.
In their view, \gls{vgi} contributors act on some combination of eight positive motivations:
(1) altruism;
(2) professional or personal interest;
(3) intellectual stimulation;
(4) protection or enhancement of a personal investment;
(5) social reward;
(6) enhanced personal reputation;
(7) self-expression;
(8) pride of place.

However, motivations of contributors are not exclusively positive. 
As \cite{williams:2004:understandingking} noted for virtual spaces, \gls{vgi} communities tend be made up of ``turbulent and shifting populations'' (p. 15).
Uncooperative and disruptive behaviour drains resources from volunteers, and generates tensions within projects \citep{wall:2007:policing}.
Amateur mapping communities are no exception, and suffer from trolling and cyber-bullying.
For example, \gls{osm} has been often disrupted by the deviant behaviour of a minority of individuals.
Project founder Steve \cite{coast:2010:enough} has directly intervened against deviance that plagues the community, advocating the `disinfection' of the project from `poisonous people' who ``drain, paralyse, slow, cause needless infighting and destroy the attention and focus of a community \ldots are wrecking the time, focus and goodwill of the majority of contributors, creating dissent out of nothing and even purposefully breaking our data'' (para. 8).

In the context of \gls{c}, \cite{coleman:2009:volunteered} have identified three negative motivations for contribution:
mischief (general destructive behaviour);
agenda (conscious purpose);
and malice and criminal intent (personal gain).
According to them, such motivations lead to a number of deviant behaviours, including: 
mass or partial deletes;
nonsense (incomprehensible information);
spam;
offensive content;
and misinformation.
Although this classification is a useful starting point, it fails to discriminate between the diverse and deep motivations of deviant expressions of \gls{v}.  
To date, no systematic analysis of vandalism occurring in the context of digital cartography has been conducted.
The next section fills this gap by investigating different types of \gls{v}.






\section{A typology of \protect\gls{v}}
\label{sec:classif}

\Gls{v} is an emergent phenomenon, and its boundaries are difficult to delineate precisely.
Unlike general digital vandalism, \gls{v} has a strong geographic component, and bears an intimate relationship with places.
The target of \gls{v} is primarily a digital artefact containing geographic information, such as a vector dataset, a spatial database, or a gazetteer.
The classic economic notion of the \emph{utility} of geographic information can function as the guiding principle to include or exclude a given action within \gls{v}.

As utility is context-sensitive and subjective, some acts can increase the utility of an artefact for one group of actors, while reducing it for another, and in such cases conflicts arise between groups with divergent views.
Hence, an act of \gls{v} \emph{intentionally reduces the utility of a geospatial artefact for the majority of the users}.
The definitional difficulties encountered when dealing with vandalism are particularly visible in \gls{c}.
Although the actor's intentionality is crucial to define \gls{v}, it is often difficult to assess it in practice, resulting in a wide gray area of acts that might be due to incompetence rather than maliciousness, or a combination of the two.
For this reason, it is very common for contributors flagged as vandals to claim that what they did was not vandalism, and controversies ensue.

In order to uncover the motivations of this phenomenon, a new qualitative analysis of real amateur mapping incidents was conducted.
This analysis focused on the \gls{v} incidents reported and discussed on the forums and mailing lists of \gls{wm} and \gls{osm}, two highly representative collaborative cartographic projects.
These incidents provided the empirical ground for the typology of \gls{v} outlined in this section, and summarised in Table \ref{table:geovandalism}.
The types of \gls{v} identified in this classification, framed in the tradition of \cite{cohen:1973:property}, are not mutually exclusive, but can co-exist in the same incidents.
For example, incidents of political \gls{v} often have a playful component.
This typology also emphasises the aspects that distinguish \gls{v} from other types of vandalism.


\gls{wm} is a commercial collaborative mapping project founded in 2006, ``aimed at marking all geographical objects in the world and providing a useful description of them.''\footurl{http://wikimapia.org/docs/About_Wikimapia}
On \gls{wm}, vandalism is defined as ``any deliberate action intended to corrupt information.'' 
\gls{wm}'s license agreement states that ``you will not post advertisements or solicitations of business;
you will not submit false information intentionally;
you will not vandalize or corrupt information on \gls{wm}.''\footurl{http://wikimapia.org/terms_reference.html}
The project's contributors can report suspicious behaviour on a dedicated forum, which currently contains 983 threads about vandalism.\footurl{http://wikimapia.org/forum}

\gls{osm} is a non-profit \gls{vgi} project started in 2004, focused on the construction and maintenance of a vector map of the entire planet.
In this context, ``vandalism is intentionally ignoring the consensus norms of the OpenStreetMap community,'' and includes copyright infringement, graffiti, use of bots, disruptive behaviour, and spamming.\footurl{http://wiki.openstreetmap.org/wiki/Vandalism}
Contributors can report suspect cases of vandalism on the project's mailing lists, where incidents are discussed and solved.\footurl{http://lists.openstreetmap.org}
The following sections define in detail each type of \gls{v}, discussing salient reported incidents.

\rowcolors{2}{gray!15}{white}
\renewcommand{\arraystretch}{1.3}
\begin{table}[t]
  \begin{tabular}{P{8em}P{7em}P{9em}P{8.5em}}
	\rowcolor{gray!40}
  	 Type & Primary motive & Typical expressions &  Physical equivalent \\
     \emph{Play \gls{v}} 		& Frustration with editing tools, boredom. & Deletions, simple geometric distortions. & Graffiti, damage to public schools and transport. \\ 
     \emph{Ideological \gls{v}} & Political communication. & Defacement of symbolic places, cyber-bullying against individuals and groups. & Political graffiti. \\ 
     \emph{Fantasy \gls{v}} 	& Self-expression, humour. & Imaginary and fictional places. & Hoaxes. \\ 
     \emph{Artistic \gls{v}} 	& Self-expression. & Complex geometries, polygon art. & Art graffiti, street art. \\ 
     \emph{Industrial \gls{v}} 	& Indirect profit. & Large scale, automated defacement of datasets & Industrial sabotage. \\ 
     \emph{Carto-spam} 			& Direct profit. & Promotional messages, unsolicited advertising. & E-mail and social networking spam. \\ 
  \end{tabular}
  \caption{Typology of \gls{v}}
  \label{table:geovandalism}
\end{table}


\subsection{Play \protect\gls{v}}

To create or modify geographic content, amateur contributors have to use software tools with specific affordances, which are often complex, unclear, or poorly documented. 
Play \gls{v} arises from \gls{hci}, as part of the playful exploration of affordances \citep{haklay:2010:interactinggeospatialtech}.
This type of vandalism is driven by frustration with the editing tools and, as pointed out in the case of school vandalism, by boredom.
Acts of play \gls{v} do not have a clearly intelligible purpose, and they are generally seen as meaningless (e.g. deletion of a river, haphazard movement of points).

\begin{figure}
\centering
\includegraphics[width=0.7\textwidth]{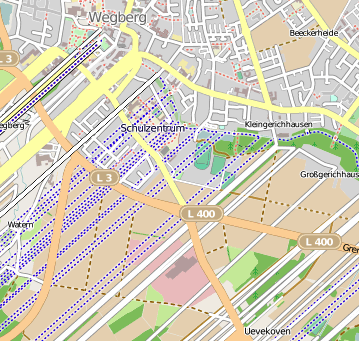}
\caption{Example of play \gls{v} resulting in damage to the street network (source: \protect\gls{osm})}
\label{fig:osm_play_van}
\end{figure}

Unclear and badly designed interfaces might increase the likelihood of contributors engaging in destructive behaviour (see Section \ref{sec:counterv}).
Play \gls{v} is generally episodic, is particularly frequent among first-time users, and is often impossible to distinguish from unintentional damage to the data. 
For example, \gls{osm} contributors using \emph{Potlatch 2} editor might not realise that they are editing the main map and not a local copy, and make exploratory changes.
Figure \ref{fig:osm_play_van} shows an incident of \gls{v} that has no precise and intelligible purpose, apart from the exploratory interaction with an editor.
Similar behaviour can be observed in videogames, in which the exploration of affordances is often accompanied by testing the destructive limits, unintended by the game designer (e.g. killing friendly soldiers or civilians in war games rather than following the game's objective).


\subsection{Ideological \protect\gls{v}}

\Gls{v} can be part of political, religious, or ethnic conflicts.
Vandalism can be driven by overtly political motives.
Electoral posters are usual targets of graffiti, through which the images of candidates are ridiculed, distorted, and defaced \citep{whalen:2012:defacingkabul}.
Urban graffiti are also deployed as markers in a conflict to indicate territorial control \citep{ley:1974:graffitimarkers}.
In these cases, the perpetrator has a clear motive, that is, sending a message to an imagined or real audience.
Acts of ideological \gls{v} include hate speech against target groups through renaming places; edits of contested borders and conflict zones; and cyberbullying against specific users.
These incidents can be understood in the framework of cyber warfare, survey by \cite{carr:2011:cyberwarfare}, in which individuals or organised groups perform attacks on digital media following a political agenda.

Military conflicts spill out to \gls{vgi} projects.
For example, in July 2012, a group of \gls{wm} users repeatedly removed and obfuscated military sites in Syria, acting within the current armed conflict between the Syrian government and the armed opposition \citep{carpenter:2013:tangledweb}.
An alleged member of the Syrian opposition added obscenity to the map, targeting the Alawites, the religious minority to which the Assad family belongs.
Another contributor edited the military academy in Aleppo, calling the Alawites `animals' in the cartographic meta-data.
The village Qardaha, home of the Assad family, was renamed as `home to monkeys.'

Similarly, during the Arab Spring, a user posted pro-government propaganda in Bahrain, while an incident of vandalism was reported in Turkey near the Syrian border.
In \gls{osm}, political \gls{v} expresses itself in the context of `tag wars,' in which different groups of contributors keep editing the same objects without reaching an agreement \citep{mooney:2012:characteristics}.
The Israeli-Palestinian conflict manifested itself in disputes over the naming of Jerusalem in the meta-data, and topoynms in the Crimea peninsula in Ukraine created controversy between Ukrainian and Russian-speaking contributors.

\subsection{Fantasy \protect\gls{v}}

Imaginary places play a key role in grounding fictional worlds in literature, art, and film \citep{joliveau:2009:realimaginaryplaces}.
More specifically, \cite{Piatti:2011:cartofictioneditorial} pointed out that ``fictional plots are set along a scale of localisations that range from the realistically rendered, highly recognisable to the completely imaginary'' (p. 218).
It should therefore not come as a surprise that the imaginations of contributors often find an outlet in creating non-existent places in \gls{c}, resulting in fantasy \gls{v}.

\begin{figure}
\centering
\includegraphics[width=0.7\textwidth]{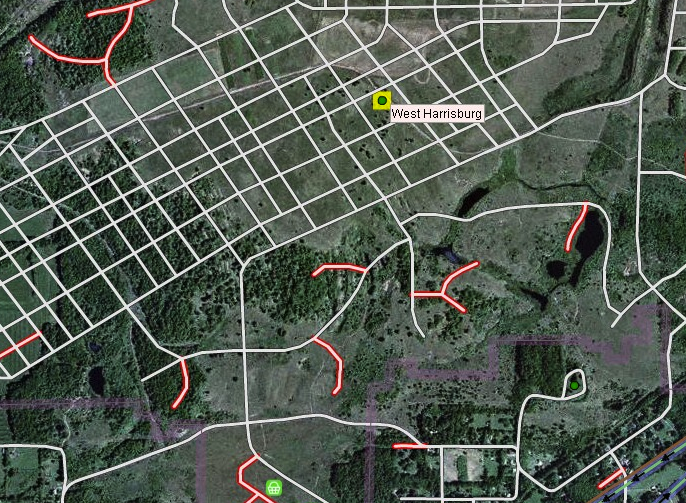}
\caption{Fantasy town in Illinois (source: \protect\gls{osm})}
\label{fig:osm_imaginary_town}
\end{figure}

Contributors conduct this kind of \gls{v} driven by the pleasure of creating imaginary natural and man-made features, playing the role of town planners and architects, similar to the player in a game such as SimCity \citep{shepherd:2009:videogamesgis}.
Several fictional towns have appeared in \gls{c}.
In the German \gls{osm}, \emph{Lummerland}, a fictional island that featured in works by author Michael Ende, was created and subsequently deleted.
Figure \ref{fig:osm_imaginary_town} shows an example of an imaginary town called `West Harrisburg,' designed by an \gls{osm} contributor.
Along similar lines, a French contributor created \emph{Parfaiteville}, i.e. the ``perfect town.''


Fictional places can be created as pranks.
In Wikipedia, Danish student Jens Roland created an article about an imaginary municipality, which survived 20 months and was translated into other languages \citep{nielsen:2012:wikipediasurvey}.
As \cite{monmonier:1996:liewithmaps} pointed out, cartographers have traditionally indulged in the practice of \emph{trap streets}, that is the creation of small imaginary map features to detect copyright infringment in other providers' maps. 
Cartographic pranks have also been documented.
For example, in the 1970s, Richard Ciacci, an employee in the public works department in Boulder, Colorado, added a Mount Richard to the official county map \citep{monmonier:1996:liewithmaps}.

Fantasy \gls{v} can also result in the renaming of famous places, when for example a \gls{wm} user renamed London and Paris with the name of his small hometown.
On \gls{osm}, Afghan university students assigned fake names (such as `Hillbilly Hameed') to streets without official names or streets that were subject of naming disputes.
Such humorous street names unintentionally ended up in the official Apple Maps \citep{moore:2013:afghanamateurs}.

\subsection{Artistic \protect\gls{v}}

The line between cartography and art has always been fine, and likewise that between art and vandalism.
Amateur mappers utilise drawing tools to create, move, and connect points, polylines, and polygons.
The drawing tools can inspire contributors to engage in creative endeavours, deviating from the objective of generating valid geographic information, and moving towards artistic \gls{v}.
The most visible and widespread form of physical vandalism is graffiti and, although most graffiti artists possess dubious merits, the `street art' of artists such as Vinchen and Banksy constitutes a notable and widely known exception \citep{lewisohn:2008:streetart}.

The idea of drawing massive shapes directly on the planet, only visible from high altitude, is explicitly present in the fascinating geoglyphs found in the Nazca desert, and in some recent architectural projects, such as the Dubai Waterfront.
Since the 1960s, the possibility of impressing art works directly onto the Earth surface has been pursued systematically in the context of `land art,' combining elements of minimalism, photography, sculpture, performance, and conceptual art \citep{tufnell:2006:landart}.
A more recent and related field is the so-called `GPS art,' in which the artist traces lines on a map by moving physically on a territory with a tracking device.\footurl{http://www.gpsdrawing.com}  


\begin{figure}[t]
\centering
\includegraphics[width=0.70\textwidth]{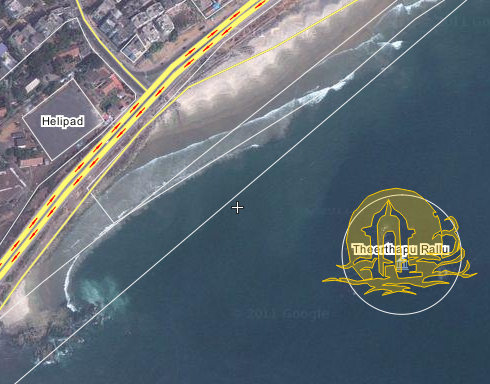}
\caption{Polygon art that depicts an underwater temple in Andhra Pradesh, India (source: \gls{wm})}
\label{fig:wm_temple}
\end{figure} 

\begin{figure}
\centering
\includegraphics[width=0.85\textwidth]{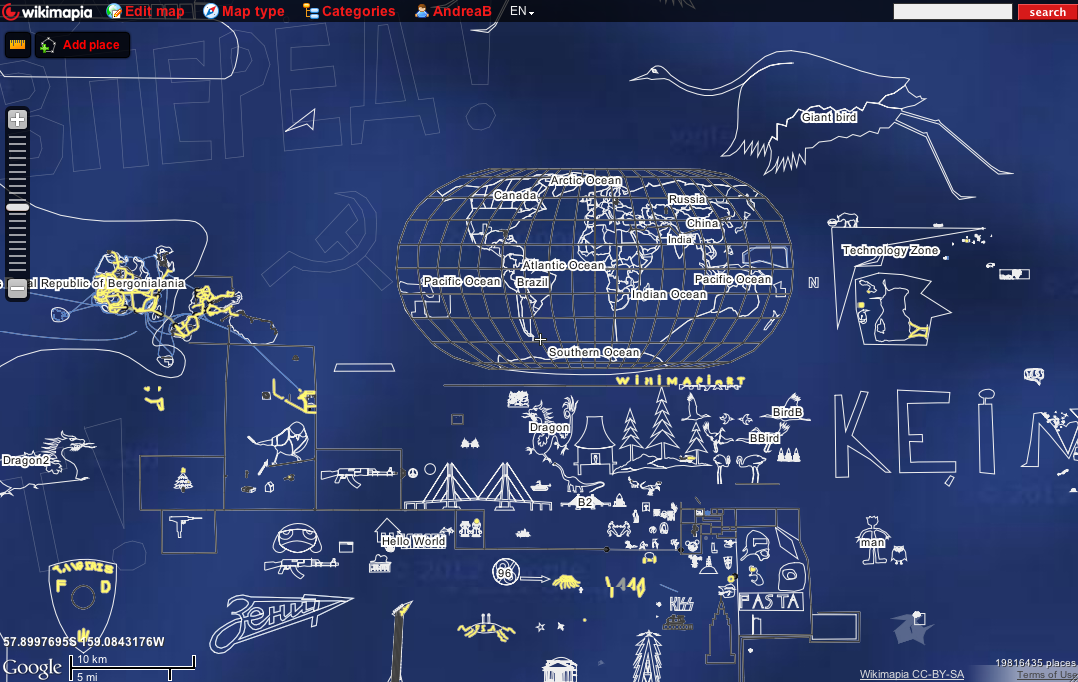}
\caption{\gls{wm}'s Polygon Art Exhibition Area, located in the Atlantic Ocean (57.899S, 159.084W)}
\label{fig:wm_polygon_art_exhib}
\end{figure}

A typical form of artistic \gls{v} is `polygon art,' the generation of aesthetically pleasing drawings using mapping tools.
Figure \ref{fig:wm_temple} shows an instance of polygon art utilised to describe an underwater temple in \gls{wm}.
Similarly to Banksy's graffiti, as \gls{wm} contributors often appreciate polygon art, an area located in the Atlantic Ocean has been created to collect valuable instances of artistic \gls{v} (see Figure \ref{fig:wm_polygon_art_exhib}).
This Polygon Art Exhibition Area acts as an art gallery, where the creative output of contributors can be safely observed without compromising the integrity of the geographic data.

\subsection{Industrial \protect\gls{v}}

Physical and digital vandalism can be performed against an organisation by one of its members, or by a hostile organisation competing in the same economic space. 
`Industrial cyber-sabotage' defines the situation where the offender damages data belonging to an organisation.
Perceived inequality and unfairness in labour relations drive disgruntled employees against their employers \citep{fisher:1982:equity}.
Another typical agent of sabotage consists of a competitor that aims at disrupting the target's economic operations, vandalising visible aspects of its business \citep{bayuk:2010:cyberforensics}.
This kind of vandalism can be carried out in a systematic way by organised groups, and is often dissimulated as other types of vandalism.
Although industrial cyber-sabotage mainly affects private corporations, allegations of incidents targeted at \gls{c} have recently been put forward.
Organisations with high economic stakes in the geo-information market could attack projects to damage their reputation and credibility as information providers.

The `Mocality affair' started in January 2012 when Mocality, a Kenyan company, reported a case of data theft carried out from machines located in the Google network infrastructure.
Subsequently, leading members of \gls{osm} reported a case of vandalism conducted from IP addresses belonging to Google \citep{maron:2012:googlevandals}.
The article stirred a raging and somewhat hysterical debate in the \gls{osm} community, partly driven by anti-corporate paranoia.
Ultimately, the incident seems to have originated from the independent initiative of low-rank contractors to Google in India (later fired), and there is no evidence of the existence of a deliberate cyber-sabotage strategy conducted by Google \citep{sottek:2010:googlevandal}.
Cartographic cyber-sabotage to date is the least likely form of \gls{v}, but is worth including as it is often discussed by amateur mapping communities.

\subsection{Cartographic spam}

As \gls{vgi} projects rely on public mailing lists, wikis, and forums, these spaces are vulnerable to traditional spam messages.
Carto-spam, by contrast, is a novel type of spam, consisting of unsolicited messages posted directly at specific locations on a map, in the features' meta-data or in the form of new features.
Driven by economic gains and generated in large volumes with the help of bots, carto-spam manifests itself on a spectrum of phenomena, some of which are obvious and some of which are more subtle.
`Obvious' carto-spam occurs in the form of messages bearing no connection with the geographic surroundings, such as hyperlinks to pornographic websites written in the meta-data of a popular building.  

At the other end of the spectrum, carto-spam is `subtle,' and consists of targeted advertisement for hotels, shops, and real estate, which bear high relevance to the geographic context.
In numerous cases, the boundary between neutral description of places and promotional messages is very thin, making its detection challenging, if not impossible.
The same issue arises in review websites, such as TripAdvisor,\footurl{http://www.tripadvisor.com} where promotional reviews are generated by the businesses' owners and skew the user ratings, in what has been called `opinion spam' by \cite{jindal:2008:opinionspam}.

In \gls{osm}, carto-spammers add specific tags to increase the visibility of businesses, for example by adding the tag \emph{tourism=attraction} to night clubs.
The emergence of carto-spam occasionally triggers debates about the scope of acceptable information, e.g. \gls{wm} contributors discussed whether ``Joe's shoe repair shop - www.joeshoefix.com'' should be considered spam or valid meta-data.
In \gls{osm}, a contributor systematically removed shop names because he did not want the project to become business-oriented.
In such cases, divergent views on the utility of geographic information determine different boundaries for \gls{v}.
What is useful for one user, can be another user's vandalism. 


\section{Countering \protect\gls{v}}
\label{sec:counterv}

The success of \gls{c} lies in its ability to harness and co-ordinate individual contributors towards the production of re-usable digital maps.
In this sense, containing and responding in a timely manner to \gls{v} is crucial to ensuring the integrity of digital artefacts, and to keeping contributors motivated. 
As in other commons-based peer production contexts, \gls{c} faces a participation/quality dilemma, a tension between the need for non-expert involvement and that of high-quality contributions.
In order to reduce the societal harm caused by physical vandalism, two approaches are adopted: repairing artefacts and prosecuting offenders in incidents that have occurred (\emph{ex-post}), and preventing vandalism from occurring (\emph{ex-ante}).
In the case of \gls{v}, legal prosecution is usually not feasible neither desirable, therefore anti-vandalism strategies focus on detecting and reverting incidents.
The only concrete way that a community can punish offenders is by banning them, which does not constitute a strong deterrent.

The issue of geo-information quality is tightly coupled with \gls{v}.
\cite{goodchild:2012:assuring} propose three families of approaches to ensure quality.
First, \emph{crowdsourcing} approaches are based on the assumption that the more users inspect an area, the higher the quality of the data.
Second, \emph{social} approaches rely on collaborative policing, adopting forms of hierarchical control, typically granting special powers to a group of selected users (e.g. moderators, administrators, etc.).
Finally, \emph{geographic} approaches aim at data validation through geographic scientific knowledge, and are to date only a theoretical possibility.
Currently, \gls{vgi} communities counter \gls{v} primarily through a combination of crowdsourcing and social approaches, setting up semi-formal mechanisms to police the vast open spaces vulnerable to malicious users, as described in the next section.


\subsection{Volunteered community involvement and policing}

To contain and manage \gls{v}, amateur mapping communities aim at identifying incidents, in what is essentially a classification problem between valid and damaging changes to an artefact.
The manual inspection of changes performed by competent contributors remains the ultimate technique to assess whether a change should be accepted or reverted.  
Hence, the social mobilisation and co-ordination of human users is key to identifying and reverting incidents.
In order to control and inhibit \gls{v}, the communities usually ban users, detecting `sockpuppets,' i.e. banned users who create new accounts to keep disrupting the project's activities.

Policing is an essential activity in the digital commons.
\cite{williams:2007:policing} observed the emergence of two policing strategies in online 3D platform Cyberworlds.
Initially, a `community involvement' model arose, in which anti-vandalism policing is performed on a non-structured basis by common users.
After a wave of highly damaging vandalism, a `volunteered community policing' model was adopted, establishing a structured response through a group of vigilantes with special powers.
This pattern occurs across commons-based peer production environments.

In Wikipedia, any user is encouraged to revert vandalous changes through community involvement, using watch lists to monitor articles of interest.
Volunteered community policing was initiated through the Counter-\-Vanda\-lism Unit, which aims to detect, fight, and to gather first-hand knowledge about vandalism in Wikipedia, with the slogan ``Civility, Maturity, Responsibility.''\footurl{http://en.wikipedia.org/wiki/Wikipedia:CVU} 
Members of the unit engage in patrolling, and are rewarded with medals of honour and badges for their efforts.
In this context, \cite{kittur:2009:herding} pointed out that belonging to a clearly defined group within the community increases the likelihood of participation in anti-vandalism and other `good citizenship' behaviours.

To counter \gls{v}, leading cartographic projects rely on volunteered community policing, coordinated through forums, wikis, and mailing lists.
In \gls{osm}, the Data Working Group deals with accusations of copyright infringement, disputes, and major cases of vandalism, indicating that ``minor incidents of vandalism should be dealt with by the local community.''\footurl{http://wiki.openstreetmap.org/wiki/Data_working_group}
Its members have the possibility of banning users only temporarily, while a restricted group of administrators have the power to ban users permanently.
The pride in one's local environment functions as an important motivational aspect, encouraging users to patrol familiar areas, tapping their local knowledge to promptly identify and revert incidents.

Being a commercial project, \gls{wm} is more centralised than \gls{osm}.
However, policing is large\-ly self-organised, and relies on gamification to motivate and reward users.
The system automatically assigns `experience points' to editors for types of actions, ranking them in different groups.
High-ranking groups, called `moderators' and `power users,' gain access to advanced map monitoring tools, fewer restrictions, and increased powers to ban deviant users.

Given their reliance on social mechanisms, such forms of policing are bound to suffer from the divides that affect \gls{vgi} in general.
As \cite{elwood:2010:gissocietal} points out, geo-information about high-in\-co\-me neighbourhoods and tourist destinations tends to be overrepresented, expressing existing divides between urban and rural, high-income and low-income social groups and areas.
Analogous coverage biases have been observed in Wikipedia \citep{george:2007:tragedywikicommons}.
Another salient divide relates to the technical skills and access to digital technologies that are needed to engage with \gls{vgi} in the first place.
In this sense, strategies to counter \gls{v} are likely to be strongly influenced by such divides, which should explicitly be taken into account.

Furthermore, policing causes social tension within the communities, especially because of the difficulties in classifying \gls{v}, and the hazy boundary between intentional and unintentional damage to the map.
Just as in any legal system, anti-vandalism patrols can abuse their banning powers, and can wrongly identify and ban a user as a vandal.
For example, a \gls{wm} user stated: ``HELP! I have been falsely banned as a clone.''
The moderator replied ``apparently you edited a few tags that were touched by a known vandal \ldots Your account name resembled a choice pattern of names the vandal had been using lately.''
To overcome these limitations, the automatic detection of \gls{v} has emerged as a complementary approach.

\subsection{Automatic detection of \protect\gls{v}}

Although human judgement is necessary in most cases to identify incidents of \gls{v}, automated procedures can provide valuable support. 
Because of the prominence of the issue and the massive size of its datasets, Wikipedia has attracted research on automatic detection of vandalism, resulting in a number of software tools and classifiers \citep{adler:2011:wikipediavandaldetect}.
After initial efforts with rule-based techniques, machine learning approaches have emerged as being more effective \cite[e.g.][]{potthast:2008:automatic}.
Indicators of Wikipedia vandalism are found in character distribution, presence of vulgar words, uppercase/lowercase ratio, semantic relatedness with the edit's context, and contributor's reputation.
Since 2010, academic competitions have been held to enhance and compare automatic vandalism detection techniques in Wikipedia, using an annotated corpus of changes as the ground truth \citep{potthast:2011:wikipediavandalsdetection}.

In the context of \gls{osm}, automatic detection is taking its first steps.
Notably, \cite{neis:2012:autovandalism} have developed \emph{OSMPatrol}, a rule-based system to detect \gls{v} in \gls{osm}.
Starting from criteria proposed by the \gls{osm} community,\footurl{http://wiki.openstreetmap.org/wiki/Detect_Vandalism} the system classifies users' actions based primarily on the contributor's reputation.
However, possibly because of an excessively inclusive definition of rules, the system seems to detect a high number of false positives, indicating edits of experienced users as potential vandalism.
\emph{OSMPatrol} confirms the difficulties of clearly discriminating between sub-optimal contributions and genuine \gls{v}.

Automatic techniques are particularly important to counter carto-spam, which is one of the most threatening forms of \gls{v} because of its for-profit motive.
In recent years, anti-spam techniques have experienced huge advances in the context of e-mails and social networking sites \citep{heymann:2007:fighting}.
However, spammers respond to new filters by re-engineering their techniques to circumvent automated barriers, resulting in a perverse feedback loop between spam and anti-spam forces.
While obvious carto-spam pushing illegal drugs and fake diplomas is easily detectable using traditional anti-spam tools, subtle carto-spam about restaurants, real estate, and tourist resorts, presents more complex challenges.
To automatically detect opinion spam, the sophisticated techniques of natural-language processing, sentiment analysis, and social network analysis discussed by \cite{jindal:2008:opinionspam} need to be combined and tailored to the cartographic domain.

\section{Conclusions}  
\label{sec:concl} 

\Gls{v} is an emergent area ripe for multi-disci\-pli\-nary research.
This article has provided a discussion of the phenomenon's salient features, motivations, and the current approaches adopted to keep it at bay with social and technological detection and control mechanisms.
Many open questions and future research directions lie ahead for cartographers, geographic information scientists, social scientists, \gls{hci} experts, and human geographers alike.

\glsreset{cpted}

  
  The typology outlined in this article was based on a direct observation of forums and mailing lists in \gls{wm} and \gls{osm}.
  To assess the impact and study the processes revolving around the generation and repression of \gls{v}, more empirical research is needed. Social network analysis and ethnographic observations can be conducted on \gls{vgi} communities, further clarifying the reasons behind \gls{v} and how communities defend the integrity of the geospatial artefacts. Such research should also be conducted beyond the most studied \gls{c} projects, including smaller projects that might provide original solutions to the problem of \gls{v}.

To enable the development of machine learning approaches and the empirical comparison between automatic techniques of \gls{v}, a more extensive corpus of real incidents should be collected, involving the affected communities. 
  To collect instances of carto-spam, the idea of `honeypots,' i.e. spaces designed to be particularly attractive to spammers, could be adapted from the context of social networking websites and applied to \gls{vgi}. 
Existing counter-\gls{v} techniques are \emph{ex-post}, aiming at identifying incidents that have already occurred.
  By contrast, \emph{ex-ante} approaches of prevention have been ignored.
  \Gls{hci} principles to design \gls{vgi} online spaces can draw on the architectural ideas developed in \gls{cpted}, which aim at designing criminogenic factors out of the built environment \citep{cozens:2008:histcpted}.
  
  Furthermore, as much play \gls{v} is caused by the interaction with complex editing tools, research in geospatial \gls{hci} is needed to identify design principles to facilitate \gls{vgi} production and monitoring, preventing, and not only repairing \gls{v} incidents.
  Relevant ideas can also be sought in the area of videogame design.
  As \cite{shepherd:2009:videogamesgis} point out, many modern videogames have interfaces to interact with complex geographic information, and provide useful ideas to improve the often unsophisticated mapping tools used by amateur mappers.

The study of \gls{v} can help understand the issue of the long-term sustainability of \gls{c}, often mentioned as one of its critical weak points \citep{dodge:2013:mappingexp}.
The geospatial digital commons are threatened by a peculiar form of the so-called `tragedy of the commons.'
The tragedy of the physical commons occurs when the behaviour of individuals, driven by their self interest, ultimately leads to the depletion of a finite resource, such as the atmosphere, the rainforest, and the reserves of fossil fuels.
As \cite{jayaraman:2012:tragedy} has suggested, the digital commons are affected not by the overexploitation that plagues the physical commons, but by \emph{underexploitation}:
the tragedy arises precisely when commons-based peer production projects lose participants and their labour, and the digital artefacts see their utility decreasing, ultimately leading to their disappearance. 
In his discussion of Wikipedia, \cite{george:2007:tragedywikicommons} argued that,  to avoid the tragedy in the long run, complex governance is needed, carefully managing vandalism and incentivising high-quality contributions.
 
This issue is of particular relevance for cartographic artefacts, which suffers from rapid obsolescence.
Every map is by definition a historical map that reflects a past state of affairs and needs updating.
To keep a map alive, constant and systematic efforts are necessary, keeping the gap between the map and the territory constant.   
It is possible to imagine that once a commons-based  project has lost its initial thrust and popular appeal, \gls{v} can damage a dataset's value irreparably, leading to the project's eventual demise.\footnote{In this sense, the British Library-funded UK Web Archive acts as a graveyard for dead websites. \url{http://www.webarchive.org.uk/ukwa}}
To date, no survey exists of the life expectancy of commons-based peer production projects.
Such research directions around \gls{v} might offer insights on the issue of maintainability, and help identify important factors in the survival of
\gls{vgi} projects amidst turbulent and unstable flows of contributions.

\bibliography{../bib/thesis,../bib/mypub} 

\begin{thebibliography}{}

\bibitem[Adler et~al., 2011]{adler:2011:wikipediavandaldetect}
Adler, B.~T., De~Alfaro, L., Mola-Velasco, S.~M., Rosso, P., and West, A.~G.
  (2011).
\newblock {Wikipedia vandalism detection: Combining natural language, metadata,
  and reputation features}.
\newblock In Gelbukh, A., editor, {\em Computational Linguistics and
  Intelligent Text Processing}, LNCS, pages 277--288. Springer.

\bibitem[Allen and Greenberger, 1978]{allen:1978:aesthetic}
Allen, V. and Greenberger, D. (1978).
\newblock {An aesthetic theory of vandalism}.
\newblock {\em Crime \& Delinquency}, 24(3):309--321.

\bibitem[Ballatore et~al., 2013]{Ballatore:2012:survey}
Ballatore, A., Wilson, D., and Bertolotto, M. (2013).
\newblock {A Survey of Volunteered Open Geo-Knowledge Bases in the Semantic
  Web}.
\newblock In Pasi, G., Bordogna, G., and Jain, L., editors, {\em Quality Issues
  in the Management of Web Information}, volume~50 of {\em Intelligent Systems
  Reference Library}, pages 93--120. Springer.

\bibitem[Bayuk, 2010]{bayuk:2010:cyberforensics}
Bayuk, J. (2010).
\newblock {\em {CyberForensics: Understanding Information Security
  Investigations}}.
\newblock Humana Press, New York.

\bibitem[Benkler and Nissenbaum, 2006]{benkler:2006:commons}
Benkler, Y. and Nissenbaum, H. (2006).
\newblock {Commons-based Peer Production and Virtue}.
\newblock {\em Journal of Political Philosophy}, 14(4):394--419.

\bibitem[Carpenter, 2013]{carpenter:2013:tangledweb}
Carpenter, T.~G. (2013).
\newblock {Tangled Web: The Syrian Civil War and Its Implications}.
\newblock {\em Mediterranean Quarterly}, 24(1):1--11.

\bibitem[Carr, 2011]{carr:2011:cyberwarfare}
Carr, J. (2011).
\newblock {\em {Inside cyber warfare: Mapping the cyber underworld}}.
\newblock O'Reilly Media, Sebastopol, CA.

\bibitem[Coast, 2010a]{coast:2010:enough}
Coast, S. (2010a).
\newblock {Enough is enough: disinfecting OSM from poisonous people (August 10,
  2010)}.
\newblock {\em OpenGeoData}.
\newblock Available from:
  \url{http://opengeodata.org/enough-is-enough-disinfecting-osm-from-poison}.

\bibitem[Coast, 2010b]{Coast:2010:bestmap}
Coast, S. (2010b).
\newblock {OpenStreetMap - The Best Map (February 19, 2010)}.
\newblock {\em OpenGeoData}.
\newblock Available from:
  \url{http://opengeodata.org/openstreetmap-the-best-map}.

\bibitem[Cohen, 1973]{cohen:1973:property}
Cohen, S. (1973).
\newblock {Property destruction: Motives and meanings}.
\newblock In Ward, C., editor, {\em Vandalism}, pages 23--53. Architectural
  Press, London.

\bibitem[Coleman et~al., 2009]{coleman:2009:volunteered}
Coleman, D., Georgiadou, Y., and Labonte, J. (2009).
\newblock {Volunteered Geographic Information: the nature and motivation of
  produsers}.
\newblock {\em International Journal of Spatial Data Infrastructures Research},
  4(2009):332--358.

\bibitem[Cozens, 2008]{cozens:2008:histcpted}
Cozens, P. (2008).
\newblock {Crime prevention through environmental design}.
\newblock In Wortley, R. and Mazerolle, L., editors, {\em Environmental
  criminology and crime analysis}, pages 153--194. Willan, Devon, UK.

\bibitem[Dodge and Kitchin, 2013]{dodge:2013:mappingexp}
Dodge, M. and Kitchin, R. (2013).
\newblock {Mapping experience: Crowdsourced cartography}.
\newblock {\em Environment and Planning A}, 45(1):19--36.

\bibitem[Elwood, 2008]{elwood:2008:volunteered}
Elwood, S. (2008).
\newblock {Volunteered geographic information: Key questions, concepts and
  methods to guide emerging research and practice}.
\newblock {\em GeoJournal}, 72(3):133--135.

\bibitem[Elwood, 2010]{elwood:2010:gissocietal}
Elwood, S. (2010).
\newblock {Geographic information science: emerging research on the societal
  implications of the geospatial web}.
\newblock {\em Progress in Human Geography}, 34(3):349--357.

\bibitem[Elwood et~al., 2012]{elwood:2012:researching}
Elwood, S., Goodchild, M., and Sui, D. (2012).
\newblock {Researching Volunteered Geographic Information: Spatial Data,
  Geographic Research, and New Social Practice}.
\newblock {\em Annals of the Association of American Geographers},
  102(3):571--590.

\bibitem[Fisher and Baron, 1982]{fisher:1982:equity}
Fisher, J. and Baron, R. (1982).
\newblock {An equity-based model of vandalism}.
\newblock {\em Population \& Environment}, 5(3):182--200.

\bibitem[Flanagin and Metzger, 2008]{Flanagin:2008:credibility}
Flanagin, A. and Metzger, M. (2008).
\newblock {The credibility of volunteered geographic information}.
\newblock {\em GeoJournal}, 72(3):137--148.

\bibitem[Furnell, 2002]{furnell:2002:cybercrime}
Furnell, S. (2002).
\newblock {Cybercrime: Vandalizing the Information Society}.
\newblock In Cueva~Lovelle, J., Gonz\'alez~Rodr\'guez, B., Joyanes~Aguilar, L.,
  Labra~Gayo, J., and del Puerto Paule~de Ruiz, M., editors, {\em Web
  Engineering}, volume 2722 of {\em LNCS}, pages 8--16. Springer.

\bibitem[Gau and Pratt, 2010]{gau:2010:revbrokenwindow}
Gau, J.~M. and Pratt, T.~C. (2010).
\newblock {Revisiting Broken Windows Theory: Examining the Sources of the
  Discriminant Validity of Perceived Disorder and Crime}.
\newblock {\em Journal of Criminal Justice}, 38(4):758--766.

\bibitem[George, 2007]{george:2007:tragedywikicommons}
George, A. (2007).
\newblock {Avoiding Tragedy in the Wiki-Commons}.
\newblock {\em Virginia Journal of Law and Technology}, 12(8):1--42.

\bibitem[Goldstein, 1996]{goldstein:1996:psychology}
Goldstein, A. (1996).
\newblock {\em {The Psychology of Vandalism}}.
\newblock Plenum Press, New York, NY.

\bibitem[Goodchild, 2007]{Goodchild:2007:citizens}
Goodchild, M. (2007).
\newblock {Citizens as Sensors: The World of Volunteered Geography}.
\newblock {\em GeoJournal}, 69(4):211--221.

\bibitem[Goodchild and Li, 2012]{goodchild:2012:assuring}
Goodchild, M. and Li, L. (2012).
\newblock {Assuring the Quality of Volunteered Geographic Information}.
\newblock {\em Spatial Statistics}, 1:110--120.

\bibitem[Graham, 2010]{graham:2010:neogeography}
Graham, M. (2010).
\newblock {Neogeography and the Palimpsests of Place: Web 2.0 and the
  Construction of a Virtual Earth}.
\newblock {\em Tijdschrift voor economische en sociale geografie},
  101(4):422--436.

\bibitem[Haklay, 2010a]{haklay:2010:good}
Haklay, M. (2010a).
\newblock {How good is volunteered geographical information? A comparative
  study of OpenStreetMap and Ordnance Survey datasets}.
\newblock {\em Environment and Planning B: Planning and Design},
  37(4):682--703.

\bibitem[Haklay, 2010b]{haklay:2010:interactinggeospatialtech}
Haklay, M., editor (2010b).
\newblock {\em Interacting with Geospatial Technologies}.
\newblock John Wiley \& Sons, Chichester, UK.

\bibitem[Heymann et~al., 2007]{heymann:2007:fighting}
Heymann, P., Koutrika, G., and Garcia-Molina, H. (2007).
\newblock {Fighting Spam on Social Web Sites: A Survey of Approaches and Future
  Challenges}.
\newblock {\em IEEE Internet Computing}, 11(6):36--45.

\bibitem[Jayaraman, 2012]{jayaraman:2012:tragedy}
Jayaraman, K. (2012).
\newblock {Tragedy of the Commons in the Production of Digital Artifacts}.
\newblock {\em International Journal of Innovation, Management and Technology},
  3(5):625--627.

\bibitem[Jindal and Liu, 2008]{jindal:2008:opinionspam}
Jindal, N. and Liu, B. (2008).
\newblock {Opinion spam and analysis}.
\newblock In {\em Proceedings of the International Conference on Web Search and
  Web Data Mining, WSDM 2008}, pages 219--230, New York. ACM.

\bibitem[Joliveau, 2009]{joliveau:2009:realimaginaryplaces}
Joliveau, T. (2009).
\newblock {Connecting real and imaginary places through geospatial
  technologies: Examples from set-jetting and art-oriented tourism}.
\newblock {\em Cartographic Journal, The}, 46(1):36--45.

\bibitem[Jurgenson, 2012]{jurgenson:2012:atomsbits}
Jurgenson, N. (2012).
\newblock {When atoms meet bits: Social media, the mobile web and augmented
  revolution}.
\newblock {\em Future Internet}, 4(1):83--91.

\bibitem[Kelling and Wilson, 1982]{kelling:1982:brokenwindow}
Kelling, G. and Wilson, J. (1982).
\newblock {Broken windows: The police and neighborhood safety}.
\newblock {\em Atlantic Monthly}, 249(3):29--38.

\bibitem[Kittur et~al., 2009]{kittur:2009:herding}
Kittur, A., Pendleton, B., and Kraut, R. (2009).
\newblock {Herding the cats: The influence of groups in coordinating peer
  production}.
\newblock In {\em Proceedings of the 5th International Symposium on Wikis and
  Open Collaboration}, pages 1--7. ACM.

\bibitem[Lewisohn, 2008]{lewisohn:2008:streetart}
Lewisohn, C. (2008).
\newblock {\em {Street art: The graffiti revolution}}.
\newblock Abrams, New York.

\bibitem[Ley and Cybriwsky, 1974]{ley:1974:graffitimarkers}
Ley, D. and Cybriwsky, R. (1974).
\newblock {Urban Graffiti as Territorial Markers}.
\newblock {\em Annals of the Association of American Geographers},
  64(4):491--505.

\bibitem[Maron et~al., 2012]{maron:2012:googlevandals}
Maron, M., Slater, G., and Coast, S. (2012).
\newblock {Google IP Vandalizing OpenStreetMap (January 16, 2012)}.
\newblock {\em OpenGeoData}.
\newblock Available from:
  \url{http://opengeodata.org/google-ip-vandalizing-openstreetmap}.

\bibitem[Minor, 1999]{minor:1999:mussolini}
Minor, H. (1999).
\newblock {Mapping Mussolini: Ritual and Cartography in Public Art during the
  Second Roman Empire}.
\newblock {\em Imago Mundi}, 51(1):147--162.

\bibitem[Monmonier, 1996]{monmonier:1996:liewithmaps}
Monmonier, M. (1996).
\newblock {\em {How to Lie With Maps}}.
\newblock University of Chicago Press, Chicago, IL.

\bibitem[Mooney and Corcoran, 2012]{mooney:2012:characteristics}
Mooney, P. and Corcoran, P. (2012).
\newblock {Characteristics of heavily edited objects in OpenStreetMap}.
\newblock {\em Future Internet}, 4(1):285--305.

\bibitem[Mooney et~al., 2010]{mooney:2010:towards}
Mooney, P., Corcoran, P., and Winstanley, A. (2010).
\newblock {Towards quality metrics for OpenStreetMap}.
\newblock In {\em Proceedings of the 18th SIGSPATIAL International Conference
  on Advances in Geographic Information Systems}, pages 514--517, New York.
  ACM.

\bibitem[Moore, 2013]{moore:2013:afghanamateurs}
Moore, U. (2013).
\newblock {How Afghan Amateur Mappers Unintentionally Punked Apple (January 14,
  2013)}.
\newblock {\em UN Dispatch}.
\newblock Available from:
  \url{http://www.undispatch.com/how-afghan-mappers-punked-apple}.

\bibitem[Moser, 1992]{moser:1992:vandalism}
Moser, G. (1992).
\newblock {What is Vandalism? Towards a Psycho-Social Definition and Its
  Implications}.
\newblock In Christensen, H.~H., Johnson, D.~R., and Brookes, M.~H., editors,
  {\em Vandalism: Research, Prevention, and Social Policy}, pages 51--59. U.S.
  Dept. of Agriculture, Forest Service, Pacific Northwest Research Station.

\bibitem[Neis et~al., 2012]{neis:2012:autovandalism}
Neis, P., Goetz, M., and Zipf, A. (2012).
\newblock {Towards Automatic Vandalism Detection in OpenStreetMap}.
\newblock {\em ISPRS International Journal of Geo-Information}, 1(3):315--332.

\bibitem[Nielsen, 2012]{nielsen:2012:wikipediasurvey}
Nielsen, F. (2012).
\newblock {Wikipedia Research and Tools: Review and Comments}.
\newblock {\em Social Science Research Network}, pages 1--52.
\newblock Available from: \url{http://ssrn.com/abstract=2129874}.

\bibitem[Piatti and Hurni, 2011]{Piatti:2011:cartofictioneditorial}
Piatti, B. and Hurni, L. (2011).
\newblock {Cartographies of Fictional Worlds}.
\newblock {\em Cartographic Journal, The}, 48(4):218--223.

\bibitem[Potthast and Holfeld, 2011]{potthast:2011:wikipediavandalsdetection}
Potthast, M. and Holfeld, T. (2011).
\newblock Overview of the 2nd international competition on wikipedia vandalism
  detection.
\newblock In {\em CLEF 2011 Labs and Workshop, Notebook Papers, 19-22 September
  2011, Amsterdam, The Netherlands}.

\bibitem[Potthast et~al., 2008]{potthast:2008:automatic}
Potthast, M., Stein, B., and Gerling, R. (2008).
\newblock {Automatic Vandalism Detection in Wikipedia}.
\newblock In {\em Advances in Information Retrieval}, volume 4956 of {\em
  LNCS}, pages 663--668. Springer.

\bibitem[Shepherd and Bleasdale-Shepherd, 2009]{shepherd:2009:videogamesgis}
Shepherd, I.~D. and Bleasdale-Shepherd, I.~D. (2009).
\newblock {Videogames: the new GIS?}
\newblock In Lin, H. and Batty, M., editors, {\em Virtual Geographic
  Environments}, pages 311--344. Science Press, Beijing, China.

\bibitem[Sottek, 2012]{sottek:2010:googlevandal}
Sottek, T. (2012).
\newblock {Sources: Google contractors fired after vandalizing OpenStreetMap,
  `no connection' with Mocality incident (January 1, 2012)}.
\newblock {\em The Verge}.
\newblock Available from:
  \url{http://www.theverge.com/2012/1/17/2714044/google-contractors-sacked-vandalism-openstreetmap}.

\bibitem[Suler, 2004]{suler:2004:disinhibition}
Suler, J. (2004).
\newblock {The Online Disinhibition Effect}.
\newblock {\em Cyberpsychology \& Behavior}, 7(3):321--326.

\bibitem[Sutton, 1987]{sutton:1987:differential}
Sutton, M. (1987).
\newblock {\em {Differential rates of vandalism in a new town: Towards a theory
  of relative place}}.
\newblock PhD thesis, Lancashire Polytechnic.

\bibitem[TomTom, 2012]{tomtom:2012:opensource}
TomTom (2012).
\newblock {Open source maps and their alternatives}.
\newblock {\em TomTom newsletter}.
\newblock Available from:
  \url{http://www.tomtom.com/en_gb/licensing/newsletter/201205/didyouknow}.

\bibitem[Tufnell, 2006]{tufnell:2006:landart}
Tufnell, B. (2006).
\newblock {\em {Land art}}.
\newblock Tate Publishing, London.

\bibitem[Wall and Williams, 2007]{wall:2007:policing}
Wall, D. and Williams, M. (2007).
\newblock {Policing diversity in the digital age: Maintaining order in virtual
  communities}.
\newblock {\em Criminology and Criminal Justice}, 7(4):391--415.

\bibitem[Whalen, 2013]{whalen:2012:defacingkabul}
Whalen, K. (2013).
\newblock {Defacing Kabul: An iconography of political campaign posters}.
\newblock {\em Cultural Geographies}, 20(4):541--549.

\bibitem[Williams, 2004]{williams:2004:understandingking}
Williams, M. (2004).
\newblock {Understanding King Punisher and His Order: Vandalism in an Online
  Community - Motives, Meanings and Possible Solutions}.
\newblock {\em Internet Journal of Criminology}.
\newblock Available from: \url{http://www.internetjournalofcriminology.com}.

\bibitem[Williams, 2006]{williams:2006:virtualcriminal}
Williams, M. (2006).
\newblock {\em {Virtually Criminal: Crime, Deviance, and Regulation Online}}.
\newblock Taylor \& Francis, Abington, UK.

\bibitem[Williams, 2007]{williams:2007:policing}
Williams, M. (2007).
\newblock {Policing and Cybersociety: The maturation of regulation within an
  online community}.
\newblock {\em Policing \& Society}, 17(1):59--82.

\bibitem[Wittel, 2001]{wittel:2001:netsociality}
Wittel, A. (2001).
\newblock {Toward a network sociality}.
\newblock {\em Theory, culture \& society}, 18(6):51--76.

\end{thebibliography}
\bibliographystyle{apalike-url}
 
\end{document}